\documentclass[prd,aps,floats,amssymb,preprint,nofootinbib]{revtex4}

\usepackage{epsfig}
\usepackage{amssymb}
\usepackage{bbm}
\usepackage{color}
\usepackage{ulem}

\usepackage{enumitem}
\usepackage{graphicx}
\usepackage{subfigure}
\usepackage{amsmath}
\usepackage{bbm}
\usepackage{color}

\newcommand{\be}{\begin{equation}}
\newcommand{\ee}{\end{equation}}
\newcommand{\bea}{\begin{eqnarray}}
\newcommand{\eea}{\end{eqnarray}}
\newcommand{\beas}{\begin{eqnarray*}}
\newcommand{\eeas}{\end{eqnarray*}}
\newcommand{\nn}{\nonumber}

\newcommand{\half}{{1\over 2}}

\newcommand\mpl{M_{\rm Pl}}

\begin{document}

%=====TITLE AND ABSTRACT===============================================================

\title{Some Adventures in the Search for a Modified Gravity Explanation for Cosmic Acceleration}
\author{Mark Trodden\footnote{trodden@physics.upenn.edu}}

\affiliation{Center for Particle Cosmology, Department of Physics and Astronomy, University of Pennsylvania,
Philadelphia, Pennsylvania 19104, USA}

\date{\today}

\begin{abstract}

The discovery of cosmic acceleration has raised the intriguing possibility that we are witnessing the first breakdown of General Relativity
on cosmological scales. In this article I will briefly review current attempts to construct a theoretically consistent and 
observationally viable modification of gravity that is capable of describing the accelerating universe. I will discuss $f(R)$ models, and 
their obvious extensions, and the DGP model as an example of extra-dimensional implementations. I will then briefly describe the
Galileon models and their very recent multifield and curved space extensions - a class of four-dimensional effective field theories encoding extra dimensional modifications to gravity. This article is dedicated to the career of my friend and former colleague, Joshua Goldberg, and is written to appear in his festschrift.

\end{abstract}

\maketitle

\setcounter{footnote}{0}

\section{Introduction}
General Relativity (GR) is one of the best tested theories ever developed. From laboratory tests, through solar system experiments and indirect gravitational wave measurements, to the broad understanding of the expansion of the universe, the theory has performed breathtakingly well. Nevertheless, in recent years, a number of researchers have revisited the question of whether gravity might hold surprises for us, not in the high curvature regime, in which the effective field theory approach tells us GR should break down, but in the opposite, long-distance limit. A primary influence behind this interest is the observed acceleration of the universe. This effect may, of course, be due to an unnaturally small cosmological constant, or a new contribution to the mass-energy of the universe (dark energy). However, a provocative possibility is that GR itself may not provide the correct set of rules with which to understand how the known matter and radiation content affects the universe on the largest scales. It may be that curvatures and length scales in the observable universe are only now reaching values at which an infrared modification of gravity can make itself apparent by driving self-acceleration (for reviews see~\cite{Copeland:2006wr,Frieman:2008sn,Silvestri:2009hh,Caldwell:2009ix,DeFelice:2010aj}).

When I began working on this idea, as an Assistant Professor at Syracuse University, I was quite surprised that one of the people responsible for our deep understanding of GR - Josh Goldberg - was so supportive of trying to tinker with this theory. In fact, Josh seemed to be positively excited at the possibility that there might be a crack in his favorite theory, and frequently discussed some of the ideas we were working on with me.

In this article I will review some of the ways in which researchers have approached the question of modifying gravity to address cosmic acceleration. I will describe the difficulties in doing this in a sensible way, and focus on those ideas that I have worked on, beginning with those that I have discussed with Josh, and moving on briefly to some more recent progress stemming from extra dimensional approaches.
Thus, I will begin by discussing $f(R)$ models, and then briefly discuss their generalizations. I will then turn to induced gravity models, beginning with the DGP model, and then describing some generalizations and the fascinating four-dimensional effective field theories to which they reduce. Since this article is intended to mostly discuss my own work, in which Josh has shown an interest, and because of space constraints, my referencing will be necessarily idiosyncratic, and reflect papers that have influenced my thinking, plus some reviews.

\section{Modifying Gravity}
The metric tensor, the basic element of General Relativity, contains, in principle, more degrees of freedom than the usual spin-2 graviton.
The reason why one doesn't hear of these degrees of freedom in GR is that the Einstein-Hilbert action is a very special choice, resulting in second-order equations of motion, which constrain away the scalars and the vectors, so that they are non-propagating. However, this is not the case if one departs from the Einstein-Hilbert form for the action. When using any modified action (and the usual variational principle) one inevitably frees up some of the additional degrees of freedom. In fact, this can be a good thing, in that the dynamics of these new degrees of freedom may be precisely what one needs to drive the accelerated expansion of the universe. However, there is often a price to pay.

The problems may be of several different kinds. First, there is the possibility that along with the desired deviations from GR on cosmological scales, one may also find similar deviations on solar system scales, at which GR is rather well-tested. Second is the possibility that the newly-activated degrees of freedom may be badly behaved in one way or another; either having the wrong sign kinetic terms (ghosts), and hence being unstable, or leading to superluminal propagation, which may lead to other problems.

These constraints are surprisingly restrictive when one tries to create viable modified gravity models yielding cosmic acceleration. 

\subsection{A Simple Model: $f(R)$ Gravity}
The simplest modification one could think of is to replace the Einstein-Hilbert Lagrangian density by a general function $f(R)$ of the Ricci scalar $R$~\cite{Carroll:2003wy,Capozziello:2003tk}.
\be
S=\frac{\mpl^2}{2}\int d^4 x\sqrt{-g}\, \left[R+f(R)\right] + \int d^4 x\sqrt{-g}\, {\cal L}_{\rm m}[\chi_i,g_{\mu\nu}] \ ,
\label{jordanaction}
\ee
where $\mpl\equiv (8\pi G)^{-1/2}$ is the (reduced) Planck mass and ${\cal L}_{\rm m}$ is the Lagrangian
density for the matter fields $\chi_i$.

Here, the matter Lagrangian is written as ${\cal L}_{\rm m}[\chi_i,g_{\mu\nu}]$ to make explicit that in
this frame - the {\it Jordan} frame - matter falls along geodesics of the metric $g_{\mu\nu}$.

The equation of motion obtained by varying the action~(\ref{jordanaction}) is
\be
\left(1+f_R \right)R_{\mu\nu} - \frac{1}{2}g_{\mu\nu}\left(R+f\right)
+ \left(g_{\mu\nu}\nabla^2 -\nabla_\mu\nabla_\nu\right) f_R
=\frac{T_{\mu\nu}}{\mpl^2} \ ,
\label{jordaneom}
\ee
where $f_R\equiv \partial f/\partial R$.

Describing the matter content as a perfect fluid, with energy-momentum tensor,
\begin{equation}
T_{\mu\nu}^m = (\rho_m + p_m)U_{\mu} U_{\nu} + p_m g_{\mu\nu}\ ,
\label{perfectfluid}
\end{equation} 
where $U^{\mu}$ is the fluid rest-frame four-velocity, $\rho_m$ is the energy density and $p_m$ is the pressure, the fluid equation of motion is then the usual continuity equation. 

When considering the background cosmological evolution of such models, the metric can be taken as the flat Robertson-Walker form, $ds^2=-dt^2+a^2(t)d{\bf x}^2$. In this case, the usual Friedmann equation of GR is modified to become
\be
3H^2 -3f_R ({\dot H}+H^2)+\frac{1}{2}f+18f_{RR}H({\ddot H}+4H{\dot H})=\frac{\rho_m}{\mpl^2}
\label{jordanfriedmann}
\ee
and the continuity equation is
\be
{\dot \rho}_m +3H(\rho_m+p_m)=0 \ .
\label{jordancontinuity}
\ee
When supplied with an equation of state parameter $w$, the above equations are sufficient to solve for the
background cosmological behavior of the space-time and it's matter contents. For appropriate choices of the
function $f(R)$ it is possible to obtain late-time cosmic acceleration without the need for dark energy, 
although evading bounds from precision solar-system tests of gravity turns out to be a much trickier
matter, as we shall see.

It is convenient to perform a carefully-chosen conformal transformation on the metric, in order to render the gravitational action in the usual Einstein Hilbert form of GR. We therefore write
\be
{\tilde g}_{\mu\nu} = \Omega(x^{\alpha}) g_{\mu\nu} \ ,
\label{conftrans}
\ee
and construct the function $r(\Omega)$ that satisfies
\be
1+f_R[r(\Omega)]=\Omega \ .
\ee
Defining a rescaled scalar field by $\Omega \equiv e^{\beta\phi}$, with 
$\beta\mpl\equiv\sqrt{2/3}$, the resulting action becomes
\bea
{\tilde S}=\frac{\mpl}{2}\int d^4 x\sqrt{-{\tilde g}}\, {\tilde R} &+&\int d^4 x\sqrt{-{\tilde g}}\, 
\left[-\frac{1}{2}{\tilde g}^{\mu\nu}(\partial_{\mu}\phi)\partial_{\nu}\phi -V(\phi)\right] \nonumber \\ 
&+&
\int d^4 x\sqrt{-{\tilde g}}\, e^{-2\beta\phi} {\cal L}_{\rm m}[\chi_i,e^{-\beta\phi}{\tilde g}_{\mu\nu}]\ ,
\label{einsteinaction}
\eea
where the potential $V(\phi)$ is determined entirely by the original form~(\ref{jordanaction}) 
of the action and is given by
\be
V(\phi)=\frac{e^{-2\beta\phi}}{2}\left\{e^{\beta\phi}r[\Omega(\phi)] - f(r[\Omega(\phi)]) \right\} \ .
\label{einsteinpotential}
\ee

The equations of motion in the Einstein frame are much more familiar than those in the Jordan frame, 
although there are some crucial subtleties. In particular, note that in general, test particles of the matter content $\chi_i$ do not freely fall along geodesics of the metric ${\tilde g}_{\mu\nu}$.

The equations of motion in this frame are those obtained by varying the action with respect to the metric ${\tilde g}_{\mu\nu}$
\be
{\tilde G}_{\mu\nu} = \frac{1}{\mpl^2}\left({\tilde T}_{\mu\nu} + T^{(\phi)}_{\mu\nu}\right) \ ,
\label{einsteineom}
\ee
with respect to the scalar field $\phi$
\be
{\tilde \nabla^2}\phi = -\frac{dV}{d\phi}(\phi) \ ,
\label{scalareom}
\ee
and with respect to the matter fields $\chi_i$, described as a perfect fluid.

Once again, we specialize to consider background cosmological evolution in this frame. The 
Einstein-frame line element can be written in familiar FRW form as
\be
ds^2 =-d{\tilde t}^2+{\tilde a}^2({\tilde t})d{\bf x}^2 \ ,
\label{einsteinFRWmetric}
\ee
where $d{\tilde t}\equiv\sqrt{\Omega}\, dt$ and ${\tilde a}(t)\equiv\sqrt{\Omega} \,a(t)$. The Einstein-frame matter energy-momentum tensor is then given by
\be
{\tilde T}_{\mu\nu}^m = ({\tilde \rho}_m + {\tilde p}_m){\tilde U}_{\mu} {\tilde U}_{\nu} + 
{\tilde p}_m {\tilde g}_{\mu\nu}\ ,
\label{einsteinperfectfluid}
\ee
where ${\tilde U}_{\mu}\equiv \sqrt{\Omega} \,U_{\mu}$, ${\tilde \rho}_m\equiv \rho_m/\Omega^2$ and 
${\tilde p}_m\equiv p_m/\Omega^2$.

Now, as I mentioned in the introduction, any modification of the Einstein-Hilbert action must, of
course, be consistent with the classic solar system tests of gravity
theory, as well as numerous other astrophysical dynamical tests. 
We have chosen the coupling constant $\mu$ to be very small, but
we have also introduced a new light degree of freedom.  As shown by
Chiba~\cite{Chiba:2003ir}, the simple model above is equivalent to a Brans-Dicke
theory with $\omega=0$ in the approximation where the potential
was neglected, and would therefore be inconsistent with solar system measurements~\cite{Bertotti:2003rm}.

To construct a realistic $f(R)$ model requires a more complicated function, with more than one
adjustable parameter in order to fit the cosmological data~\cite{Bean:2006up} and satisfy solar system bounds through the
chameleon~\cite{Khoury:2003aq} mechanism. 

\subsection{Extensions: Higher-Order Curvature Invariants}
It is natural to consider generalizing the action of~\cite{Carroll:2003wy} to include other curvature invariants~\cite{Carroll:2004de}.
There are, of course, any number of terms that one could consider, but for simplicity, focus on
those invariants of lowest mass dimension that are also parity-conserving $P \equiv  R_{\mu\nu}\,R^{\mu\nu}$ and 
$Q \equiv  R_{\alpha\beta\gamma\delta}\,R^{\alpha\beta\gamma\delta}$.

The action then takes the form
\begin{equation}
S=\int d^4x \sqrt{-g}\,[R+f(R,P,Q)] +\int d^4 x\, \sqrt{-g}\,
{\cal L}_M \ ,
\label{genaction}
\end{equation}
where $f(R,P,Q)$ is a general function describing deviations from general relativity.

Actions of the form~(\ref{genaction}) generically admit a maximally-symmetric solution that is often unstable to another accelerating power-law attractor. 
It has been shown that solar system constraints, of the type I have described for $f(R)$ models, can be evaded by these more general 
models when, for example, the $Q$ terms are relevant on those scales. However, these theories generically contain ghosts and/or superluminally propagating modes~\cite{Chiba:2005nz,DeFelice:2006pg,Calcagni:2006ye}. I therefore will not discuss them further here.

\subsection{Induced Gravity Models}

In the Dvali-Gabadadze-Porrati (DGP) model~\cite{Dvali:2000hr}, our observed $4D$ universe is embedded in an infinite empty fifth dimension. 
Despite the fact that the extra dimension is infinite in extent, the inverse-square law is nevertheless recovered at short distances on the brane due to an intrinsic, four-dimensional Einstein-Hilbert term in the action
\be
	S_{\rm DGP} = \int_{\rm bulk} {\rm d}^5x\sqrt{-g_5}\frac{M_5^3}{2}R_5 +
\int_{\rm brane} {\rm d}^4x \sqrt{-g_4} \left(\frac{M_4^2}{2}R_4 + {\cal L}_{\rm matter}\right)\,.
\ee
The Newtonian potential on the brane scales as $1/r$ at short distances, as in $4D$ gravity, and asymptotes to $1/r^2$ at large distances, characteristic of $5D$ gravity. The cross-over scale $m_5^{-1}$ between these two behaviors is set by the bulk and brane Planck masses ($M_{5}$ and $M_{4}$ respectively) via $m_5 = \frac{M_5^3}{M_4^2}$.

In this picture, the higher-dimensional nature of gravity affects the $4D$ brane through deviations from general relativity on horizon scales, that may give rise to the observed accelerated expansion. This model faces its own challenges however. The branch of solutions that include self-acceleration suffers from ghost-like instabilities, and on the observational front, DGP cosmology is statistically disfavored in comparison to $\Lambda$CDM and is significantly discordant with constraints on the curvature of the universe. 

\section{Galileons}
Careful studies of the DGP model have, however, given rise to new ideas about how to construct four-dimensional effective field theories with symmetries that may be relevant for cosmology. The decoupling limit of DGP consists of a $4$-dimensional effective theory of gravity coupled to a single scalar field $\pi$, representing the bending mode of the brane in the fifth dimension. The $\pi$ field self-interaction includes a cubic self-interaction $\sim (\partial\pi)^2\square\pi$, which has the properties that the field equations are second order, and the terms are invariant up to a total derivative under the internal galilean transformations 
\be
\pi\rightarrow \pi+c+b_\mu x^\mu \ ,
\ee
where $c,b_\mu$ are arbitrary real constants.

In \cite{Nicolis:2008in}, this was generalized, and all possible lagrangian terms for a single scalar with these two properties were classified in all dimensions.  They are called galileon terms, and there exists a single galileon lagrangian at each order in $\pi$, where ``order" refers to the number of copies of $\pi$ that appear in the term.   
For $n\geq 1$, the $(n+1)$-th order galileon lagrangian is
\be
\label{galileon2} 
{\cal L}_{n+1}=n\eta^{\mu_1\nu_1\mu_2\nu_2\cdots\mu_n\nu_n}\left( \partial_{\mu_1}\pi\partial_{\nu_1}\pi\partial_{\mu_2}\partial_{\nu_2}\pi\cdots\partial_{\mu_n}\partial_{\nu_n}\pi\right),
\ee 
where 
\be
\label{tensor} 
\eta^{\mu_1\nu_1\mu_2\nu_2\cdots\mu_n\nu_n}\equiv{1\over n!}\sum_p\left(-1\right)^{p}\eta^{\mu_1p(\nu_1)}\eta^{\mu_2p(\nu_2)}\cdots\eta^{\mu_np(\nu_n)} \ .
\ee 
The sum in~(\ref{tensor}) is over all permutations of the $\nu$ indices, with $(-1)^p$ the sign of the permutation.  The tensor~(\ref{tensor}) is anti-symmetric in the $\mu$ indices, anti-symmetric the $\nu$ indices, and symmetric under interchange of any $\mu,\nu$ pair with any other.  These lagrangians are unique up to total derivatives and overall constants.   Because of the anti-symmetry requirement on $\eta$, only the first $n$ of these galileons are non-trivial in $n$-dimensions.  In addition, the tadpole term, $\pi$, is galilean invariant, and we therefore include it as the first-order galileon.  

Thus, at the first few orders, we have 
\bea
 {\cal L}_1&=&\pi, \\ \nn
 {\cal L}_2&=&[\pi^2], \\ \nn
{\cal L}_3&=&[\pi^2][\Pi]-[\pi^3], \\ \nn
{\cal L}_4&=&\half[\pi^2][\Pi]^2-[\pi^3][\Pi]+[\pi^4]-\half[\pi^2][\Pi^2], \\ \nn
{\cal L}_5&=&{1\over 6}[\pi^2][\Pi]^3-{1\over 2}[\pi^3][\Pi]^2+[\pi^4][\Pi]-[\pi^5]+{1\over 3}[\pi^2][\Pi^3]-{1\over 2}[\pi^2][\Pi][\Pi^2]+{1\over 2}[\pi^3][\Pi^2]
\ .
\eea
We have used the notation $\Pi$ for the matrix of partials $\Pi_{\mu\nu}\equiv\partial_{\mu}\partial_\nu\pi$, and $[\Pi^n]\equiv Tr(\Pi^n)$, e.g. $[\Pi]=\square\pi$, $[\Pi^2]=\partial_\mu\partial_\nu\pi\partial^\mu\partial^\nu\pi$, and $[\pi^n]\equiv \partial\pi\cdot\Pi^{n-2}\cdot\partial\pi$, i.e. $[\pi^2]=\partial_\mu\pi\partial^\mu\pi$, $[\pi^3]=\partial_\mu\pi\partial^\mu\partial^\nu\pi\partial_\nu\pi$.  The above terms are the only ones which are non-vanishing in four dimensions.  The second is the standard kinetic term for a scalar, while the third is the DGP $\pi$-lagrangian (up to a total derivative).

The first few orders of the equations of motion are 
\bea{\cal E}_1&=&1, \\
 {\cal E}_2&=&-2[\Pi], \\
{\cal E}_3&=& -3\left([\Pi]^2-[\Pi^2]\right),  \\
{\cal E}_4&=& -2\left([\Pi]^3+2[\Pi^3]-3[\Pi][\Pi^2]\right),  \\
{\cal E}_5&=& -{5\over 6}\left([\Pi]^4-6[\Pi^4]+8[\Pi][\Pi^3]-6[\Pi]^2[\Pi^2]+3[\Pi^2]^2\right) \ .
\eea

These galileon actions can be generalized to the multi-field case, where there is a multiplet $\pi^I$ of fields~\cite{Deffayet:2009mn,Deffayet:2010zh,Padilla:2010de,Hinterbichler:2010xn}.  The action in this case can be written
\be 
\label{generaltermspre} 
{\cal L}_{n+1}= S_{I_1I_2\cdots I_{n+1}}\eta^{\mu_1\nu_1\mu_2\nu_2\cdots\mu_n\nu_n}\left(\pi^{I_{n+1}} \partial_{\mu_1}\partial_{\nu_1}\pi^{I_1}\partial_{\mu_2}\partial_{\nu_2}\pi^{I_2}\cdots\partial_{\mu_n}\partial_{\nu_n}\pi^{I_{n}}\right),
\ee
with $S_{I_1I_2\cdots I_{n+1}}$ a symmetric constant tensor.  This is invariant under under individual galilean transformations for each field, $\pi^I\rightarrow \pi^I+c^I+b^I_\mu x^\mu$, and the equations of motion are second order,
\be 
{\cal E}_{I}\equiv {\delta{\cal L}\over \delta \pi^I}=(n+1)S_{II_1I_2\cdots I_{n}}\eta^{\mu_1\nu_1\mu_2\nu_2\cdots\mu_n\nu_n}\left(\partial_{\mu_1}\partial_{\nu_1}\pi^{I_1}\partial_{\mu_2}\partial_{\nu_2}\pi^{I_2}\cdots\partial_{\mu_n}\partial_{\nu_n}\pi^{I_{n}}\right) \ .
\ee
These extended models have their own constraints~\cite{Andrews:2010km}, which I won't have time to discuss further here.

The theory containing these galilean-invariant operators is not renormalizable, i.e. it is an effective field theory with a cutoff $\Lambda$, above which some UV completion is required.  However, crucially, the ${\cal L}_n$ terms above do not get renormalized upon loop corrections, so that their classical values can be trusted quantum-mechanically

In principle, one should consider quantum effects within the effective theory, since there are other operators of the same dimension that might compete with the galileon terms.  However, fascinatingly, there can exist interesting regimes where non-linearities from the galileon terms are important, yet quantum effects are under control. This separation of scales allows for the existence of regimes in which there exist classical field configurations with non-linearities of order one, and still trust these solutions in light of quantum corrections. These non-linear, quantum-controlled regimes are where interesting models of inflation, cosmology, modified gravity, etc. employing these galileon actions should be placed.

\subsection{Galileons on Curved Spaces}

While various authors have considered the uses of these theories for early and late cosmology, it would be interesting to find examples of theories with the same attractive symmetry features, but which naturally live on the curved manifolds relevant to cosmology. There are obstacles to performing this in a straightforward manner~\cite{Deffayet:2009wt}. However, in very recent
papers~\cite{Goon:2011qf,Goon:2011uw} we have demonstrated how to construct such models\footnote{This construction has connections to some other recently proposed ones~\cite{Deffayet:2011gz,Burrage:2011bt}}, and I will finish up this article by briefly describing the construction.

The general context is the theory of a dynamical 3-brane moving in a fixed (4+1)-dimensional background.  
The dynamical variables are the brane embedding $X^A(x)$, five functions of the world-volume coordinates $x^\mu$.  The bulk has a fixed background metric $G_{AB}(X)$, from which we may construct the induced metric $\bar g_{\mu\nu}(x)$ and the extrinsic curvature $K_{\mu\nu}(x)$
\bea 
\bar g_{\mu\nu}&=&e^A_{\ \mu}e^B_{\ \nu} G_{AB}(X) \ , \\ 
K_{\mu\nu}&=&e^A_{\ \mu}e^B_{\ \nu}\nabla_A n_B \ ,
\eea
where $e^A_{\ \mu}= {\partial X^A\over\partial x^\mu}$ are the tangent vectors to the brane, and $n^A$ is the unit normalized normal vector.

The world-volume action must be gauge invariant under reparametrizations of the brane, which is guaranteed if the action is written as a diffeomorphism scalar, $F$, of $\bar g_{\mu\nu}$, $K_{\mu\nu}$, the covariant derivative $\bar\nabla_\mu$ and the curvature $\bar R^\alpha_{\ \beta\mu\nu}$ constructed from $\bar g_{\mu\nu}$,
\be
\label{generalaction} 
S= \int d^4x\ \sqrt{-\bar g}F\left(\bar g_{\mu\nu},\bar\nabla_\mu,\bar R^{\alpha}_{\ \beta\mu\nu},K_{\mu\nu}\right) \ .
\ee

This action possesses global symmetries only if the bulk metric possesses Killing symmetries. We fix all the gauge symmetry by choosing the gauge
\be
\label{physgauge} 
X^\mu(x)=x^\mu, \ \ \ X^5(x)\equiv \pi(x) \ ,
\ee
where we have foliated the bulk by time-like slices given by the surfaces $X^5(x)= {\rm constant}$.  The remaining coordinates $X^\mu$ can then be chosen arbitrarily and parametrize the leaves of the foliation. In this gauge, the coordinate $\pi(x)$ measures the transverse position of the brane relative to the foliation, and the resulting action solely describes $\pi$,
\be
\label{gaugefixedaction} 
S= \int d^4x\ \left. \sqrt{-\bar g}F\left(\bar g_{\mu\nu},\bar\nabla_\mu,\bar R^{\alpha}_{\ \beta\mu\nu},K_{\mu\nu}\right)\right|_{X^\mu=x^\mu,\ X^5=\pi} \ .
\ee

Since gauge fixing cannot alter global symmetries, any global symmetries of~(\ref{generalaction}) become global symmetries of~(\ref{gaugefixedaction}).   However, the form of the global symmetries depends on the gauge because the gauge choice is not generally preserved by the global symmetry.  Given a transformation generated by a Killing vector, $K^A$, we restore our preferred gauge (\ref{physgauge}) by making a compensating gauge transformation $\delta_{g,{\rm comp}}x^\mu=-K^\mu$.  The two symmetries then combine to shift $\pi$ by
\be
\label{gaugefixsym} 
(\delta_K+\delta_{g,{\rm comp}})\pi=-K^\mu(x,\pi)\partial_\mu\pi+K^5(x,\pi) \ ,
\ee
which is a symmetry of the gauge fixed action~(\ref{gaugefixedaction}).

%%%%%%%%

It is convenient at this stage to make two simplifying assumptions. We specialize to the case in which the foliation is Gaussian normal with respect to the metric $G_{AB}$, and we demand that the extrinsic curvature on each of the slices be proportional to the induced metric. Under these assumptions the metric takes the form
\be 
\label{metricform} 
G_{AB}dX^AdX^B=d\rho^2+f(\rho)^2g_{\mu\nu}(x)dx^\mu dx^\nu \ ,
\ee
where $X^5=\rho$ denotes the transverse coordinate, and $g_{\mu\nu}(x)$ is an arbitrary brane metric.  This special case includes all examples in which a maximally symmetric ambient space is foliated by maximally symmetric slices. 

In the gauge (\ref{physgauge}), the induced metric is $\bar g_{\mu\nu}=f(\pi)^2g_{\mu\nu}+\nabla_\mu\pi\nabla_\nu\pi$.
Defining the quantity $\gamma=1/ \sqrt{1+{1\over f^2}(\nabla\pi)^2}$,
the extrinsic curvature is then
\be 
K_{\mu\nu}=\gamma\left(-\nabla_\mu\nabla_\nu\pi+f f'g_{\mu\nu}+2{f'\over f}\nabla_\mu\pi\nabla_\nu\pi\right) \ .
\ee

A general choice for the action~(\ref{gaugefixedaction}) will lead to scalar field equations for $\pi$ which are higher than second order in derivatives and may therefore propagate extra ghost degrees of freedom.  
However, as pointed out in~\cite{deRham:2010eu}, there are a finite number of actions that lead to second order equations.
The possible
extensions of Einstein gravity which remain second order are given by Lovelock terms and their boundary terms.  
These terms are specific combinations of powers of the Riemann tensor which are topological (i.e. total derivatives) in some specific home dimension, but in lower dimensions have the property that equations of motions derived from them are second order.

The prescription of~\cite{deRham:2010eu} is then as follows: on the 4-dimensional brane, we may add the first two Lovelock terms, namely the cosmological constant term $\sim \sqrt{-\bar g}$ and the Einstein-Hilbert term $\sim \sqrt{-\bar g}R[\bar g]$.  (The higher Lovelock terms will be total derivatives in 4-dimensions.)  We may also add the boundary term corresponding to a bulk Einstein-Hilbert term, $\sqrt{-\bar g}K$, and the boundary term ${\cal L}_{\rm GB}$ corresponding to the Gauss-Bonnet Lovelock invariant $R^2 - 4 R_{\mu\nu} R^{\mu\nu}+ R_{\mu\nu\alpha\beta} R^{\mu\nu\alpha\beta}$ in the bulk.  The zero order cosmological constant Lovelock term in the bulk has no boundary term, although we may construct a tadpole-like term from it, and the higher order bulk Lovelock terms vanish identically.  Therefore, in total, for a 3-brane there are five terms that lead to second order equations for $\pi$,
\bea   {\cal L}_1&=&\sqrt{-g}\int^\pi d\pi' f(\pi')^4,\nn\\
{\cal L}_2&=&- \sqrt{-\bar g} \ ,\nn\\
{\cal L}_3&=& \sqrt{-\bar g}K \ ,\nn\\
{\cal L}_4&=& -\sqrt{-\bar g}\bar R \ ,\nn\\
{\cal L}_5&=&{3\over 2}\sqrt{-\bar g} {\cal K}_{\rm GB} \ ,
\label{ghostfreegenterms} \eea
where the explicit form of the Gauss-Bonnet boundary term is
${\cal K}_{\rm GB}=-{1\over3}K^3+K_{\mu\nu}^2K-{2\over 3}K_{\mu\nu}^3-2\left(\bar R_{\mu\nu}-\half \bar R \bar g_{\mu\nu}\right)K^{\mu\nu}$. 

$\mathcal L_1$ is the zero derivative tadpole term which is the proper volume between any $\rho=$ constant surface and the brane position, $\pi(x)$.  While different in origin from the other terms, it too has the symmetry~(\ref{gaugefixsym}).  Each of these terms may appear in a general Lagrangian with an arbitrary coefficient. 

Evaluating these expressions for the metric~(\ref{metricform}) involves a lengthy calculation which ultimately yields
\bea   
{\cal L}_1&=&\sqrt{-g}\int^\pi d\pi' f(\pi')^4,\nn\\
{\cal L}_2&=&-\sqrt{-g}f^4\sqrt{1+{1\over f^2}(\partial\pi)^2},\nn\\
{\cal L}_3&=&\sqrt{-g}\left[f^3f'(5-\gamma^2)-f^2[\Pi]+\gamma^2[\pi^3]\right],\nn \\
{\cal L}_4&=& -\sqrt{-g}\Big\{{1\over\gamma}f^2R-2{\gamma}R_{\mu\nu}\nabla^\mu\pi\nabla^\nu\pi \nn\\
&&+\gamma\left[[\Pi]^2-[\Pi^2]+2{\gamma^2\over f^2}\left(-[\Pi][\pi^3]+[\pi^4]\right)\right]\nn\\
&&\!+6{f^3f''\over \gamma}\left(-1+\gamma^2\right) \!	 \nn \\
&&+2\gamma ff'\left[-4[\Pi]+{\gamma^2\over f^2}\left(f^2[\Pi]+4[\pi^3]\right)\right]\nn\\
&&-6{f^2f'^2\over \gamma}\left(1-2\gamma^2+\gamma^4\right) \Big\}.
\eea
The expression for ${\cal L}_5$ is lengthy, and can be found in~\cite{Goon:2011qf}.

In these expressions, all curvatures and covariant derivatives are those of the background metric $g_{\mu\nu}$.  The notation is as earlier, but with covariant derivatives replacing partial ones. Note that no integrations by parts have been performed in obtaining these terms.  The equations of motion contain no more than two derivatives on each field, ensuring that no extra degrees of freedom propagate around any background.

To develop the analogues of the original Galileon theory, we expand the Lagrangians in powers of $\lambda$ around some constant background, $\pi\rightarrow\pi_0+\lambda\pi$.  One can find appropriate linear combinations of the Lagrangians, $\bar{\mathcal L}_n=c_1\mathcal L_1+\ldots +c_n\mathcal L_n$, for which all terms $\mathcal{O}\left (\lambda^{n-1}\right )$ or lower are total derivatives.  This was performed in~\cite{deRham:2010eu} for the $M_4$ in $M_5$ case and the results precisely reproduce the Galileon and conformal Galileon theories, respectively. These examples are two of the six cases where both the bulk and the leaves of the foliation are maximally symmetric, and the bulk metric has a single time component. 

When this prescription is carried out for the remaining four maximally symmetric cases in which the 4d background is curved, new classes of theories are produced.  After canonical normalization, $\hat\pi={1\over L^2}\pi$ where $L$ is the $dS_4$ or $AdS_4$ radius, the Lagrangians become
\begin{eqnarray} 
\hat{\cal L}_1&=&\sqrt{-g}\hat\pi \ , \nn\\
\hat{\cal L}_2&=&-\half\sqrt{-g} \left((\partial\hat\pi)^2-{R\over 3}\hat \pi^2\right) \ ,\nn \\
\hat{\cal L}_3&=& \sqrt{-g}\left(-{1\over 2}(\partial\hat\pi)^2[\hat\Pi]-{R\over 4} (\partial\hat\pi)^2\hat\pi+{R^2\over 36}\hat\pi^3\right) \ ,\nn\\
\hat{\cal L}_4&=&\sqrt{-g}\Big[-\half(\partial\hat\pi)^2\Big([\hat\Pi]^2-[\hat\Pi^2]+{R\over 24}(\partial\hat\pi)^2\nn\\
&&+{R\over 2}\hat\pi[\hat\Pi]+{R^2\over 8}\hat\pi^2\Big)+{R^3\over 288}\hat\pi^4\Big] \ ,
\label{singlesetGalileons} 
\eea 
with ${\cal L}_5$ again found in~\cite{Goon:2011qf}.

Here $R=\pm{12\over L^2}$ is the Ricci curvature of the $dS_4$ or $AdS_4$ background. These simpler Lagrangians are Galileons that live on curved space yet retain the same number of symmetries as the full theory, whose form comes from expanding~(\ref{gaugefixsym}) in appropriate powers of $\lambda$.  In the case of a $dS_4$ background in conformal inflationary coordinates $(u,y^i)$, the non-linear symmetries are
\be \label{dSGalileontrans}
\delta_{+}\hat\pi={1\over u}\left(u^2-y^2\right) , \ \ \
\delta_{-} \hat\pi=-{1\over u},\ \ \ 
\delta_{i} \hat\pi = {y_i\over u} \ .
\ee

A striking feature of these fully covariant models which is not present in the flat space Galileon theories is the presence of potentials whose couplings are determined by the symmetries~(\ref{gaugefixsym}).  In particular, the scalar field acquires a mass of order the $dS_4$ or $AdS_4$ radius, with a value protected by the symmetries (\ref{dSGalileontrans}), so the small masses should be protected against renormalization.

\section{Conclusions} 

Among the possible explanations for the observed accelerated expansion of the universe, the possibility that general relativity may become modified on the largest scales is a particularly intriguing one. In this talk I have outlined a number of modern approaches to this problem, focusing, as expected, on those that I have been involved with in one way or another. I have described how the combined constraints of theoretical consistency, solar system measurements, and cosmological observations tightly bound the possible viable models. 

From higher dimensional constructions, such as the DGP model, an interesting set of four dimensional effective field theories - the {\it galileons} - arises, encapsulating the effects of modifying gravity. I have described how we have generalized that work to multi-galileon theories, and to covariant galileons propagating on curved backgrounds. This very recent work opens up the possibility of galileons with potentials protected by the symmetries inherited from the higher-dimensional bulk, and therefore naturally suited to inflation or describing late time cosmic acceleration.

My interest in the general area of modifying gravity as a possible explanation for cosmic acceleration began quite a few years ago. I was fortunate to be encouraged by Josh Goldberg from the very beginning, and to spend eight years as his colleague. It is a pleasure to be able to devote this article to his career.

\acknowledgments
I would like to thank Ed Glass and David Robinson for inviting me to contribute to this volume.  I would also like to thank my collaborators, from whose joint work with me I have borrowed liberally in putting together this summary. This work is supported in part by NASA ATP grant NNX08AH27G, NSF grant PHY-0930521, Department of Energy grant DE-FG05-95ER40893-A020, and by the Fay R. and Eugene L. Langberg chair.

\end{document}